\documentclass[10pt,letterpaper]{article}
\usepackage{opex3}

\begin{document}

\title{Measurement of throughput variation across a large format
volume-phase holographic grating}

\author{Naoyuki Tamura, Graham J. Murray, Ray M. Sharples, David
J. Robertson, \& Jeremy R. Allington-Smith}

\address{Centre for Advanced Instrumentation, Department of Physics,
University of Durham, South Road, Durham, DH1 3LE, UK}

\email{naoyuki.tamura@durham.ac.uk}

\begin{abstract}
In this paper, we report measurements of diffraction efficiency and
angular dispersion for a large format ($\sim$ 25 cm diameter)
Volume-Phase Holographic (VPH) grating optimized for near-infrared
wavelengths (0.9 $\sim$ 1.8 $\mu$m). The aim of this experiment is to
see whether optical characteristics vary significantly across the
grating. We sampled three positions in the grating aperture with a
separation of 5 cm between each. A 2 cm diameter beam is used to
illuminate the grating. At each position, throughput and diffraction
angle were measured at several wavelengths. It is found that whilst the
relationship between diffraction angle and wavelength is nearly the same
at the three positions, the throughputs vary by up to $\sim$ 10\% from
position to position. We explore the origin of the throughput variation
by comparing the data with predictions from coupled-wave analysis. We
find that it can be explained by a combination of small variations over
the grating aperture in gelatin depth and/or refractive index modulation
amplitude, and amount of energy loss by internal absorption and/or
surface reflection.
\end{abstract}


\section{Introduction}

Volume-Phase Holographic (VPH) gratings potentially have many advantages
over classical surface-relief gratings (\cite{barden1};
\cite{barden2}). They are already in operation in some existing
astronomical spectrographs and their use is also planned for a number of
forthcoming instruments (e.g., \cite{smith}). In applications to
spectrographs for extremely large telescopes, one has to consider that
the beam diameter in a spectrograph will become very large ($\sim$ 30 cm
or even larger) to obtain a reasonably high resolving power and
consequently large optics and dispersing elements will be demanded. In
this respect, VPH gratings may have an advantage. Unlike classical
surface relief gratings, VPH gratings with such large formats can be
rather easily fabricated at a reasonable cost. However, the optical
uniformity across the aperture has rarely been investigated, despite the
fact that during certain stages of fabrication, there is a significant
risk of introducing spatial variation. In particular, the uniformity of
diffraction efficiency and angular dispersion, the key to the
performance of a spectrograph, should be confirmed by experiments. 
We note that throughput measurements at the two positions (centre and
corner) of a VPH grating with the size of 16 cm $\times$ 20 cm and the
line density of 1520 lines/mm are presented (\cite{rallison}) but the
grating was made before the manufacturer's fabrication process was
improved and it shows a large difference in diffraction efficiency
between the centre and corner of the grating.

\begin{figure}[t]
\begin{center}
 \includegraphics[height=10cm,angle=-90,keepaspectratio]{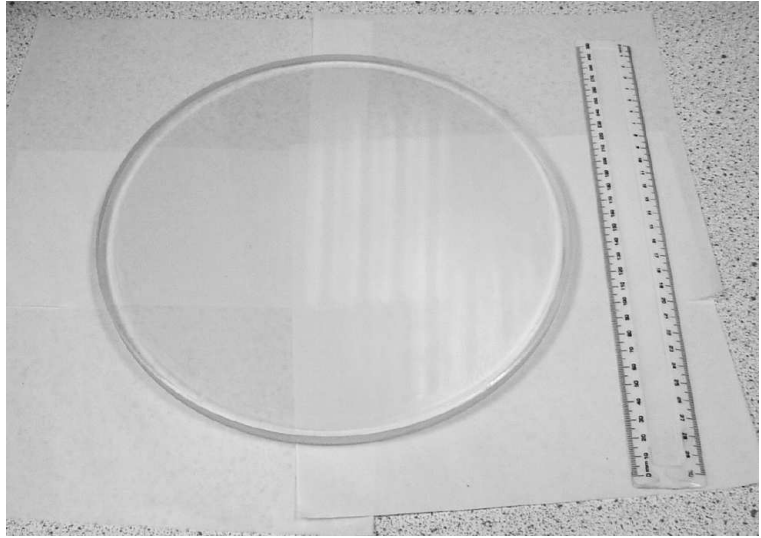}
 \caption{A picture of the sample VPH grating. The grating has a
 diameter of 250 mm with a line density of 385 lines/mm.}
 \label{vph}
\end{center}
\end{figure}

In this paper, results from measurements of throughput and angular
dispersion of a large format grating are presented. A picture of the
grating is shown in Fig. \ref{vph}. This grating was manufactured by
Ralcon Development Lab for the FMOS project (e.g., \cite{kimura}) with a
diameter of $\sim$ 25 cm. The line density is 385 lines/mm and thus the
peak of diffraction efficiency is around 1.3 $\mu$m at the Bragg
condition when the incident angle of an input beam to the normal of the
grating surface is 15$^{\circ}$. The measurements are performed at
wavelengths between 0.9 $\mu$m and 1.6 $\mu$m. In the next section, we
will describe the test facility and the measurement procedures. Note
that the grating investigated here has the same specification as that
used for the cryogenic tests (\cite{tamura}), but this element has not
previously been exposed to vacuum or cryogenic temperature.

\section{The test set-up and measurements}

\begin{figure}[t]
\begin{center}
 \includegraphics[height=8cm,angle=-90,keepaspectratio]{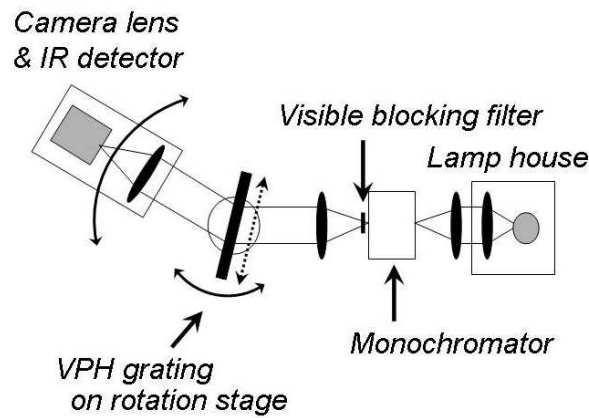}
 \caption{Schematic view of the test optics configuration.}
 \label{config}

\end{center}
\end{figure}

\setlength{\tabcolsep}{1.7mm}
\begin{table}[b]
\caption{The main components used for the measurements.}
\begin{tabular}{llll} \hline
 & Manufacturer & Product ID & \multicolumn{1}{c}{Comments} \\ \hline
Light source            & Comar             & 12 LU 100        & 
Tungsten-halogen lamp \\
Monochromator           & Oriel Instruments & Cornerstone 130, & 
600 lines/mm grating, \\
                        &                   & Model 74000      &
Blaze at 1 $\mu$m \\
Visible blocking filter & Comar             & 715 GY 50        &
 Transparent at $\lambda \geq 715$ nm \\
Near-infrared detector  & Indigo Systems    & Alpha$-$NIR      &
 320 $\times$ 256 InGaAs array \\ \hline
\end{tabular}
\label{comps}
\end{table}

In Fig. \ref{config}, the overall configuration of the optical
components used for the measurements is indicated (detailed information
for the main components is given in Table \ref{comps}).  Light exiting
from the monochromator is collimated and used as an input beam to
illuminate the VPH grating. The spectral band-width of this input beam
is set by adjusting the width of the output slit of the monochromator.
The slit width and the corresponding spectral band-width were set to 0.5
mm and $\sim$ 0.01 $\mu$m, respectively, throughout the measurements;
the beam diameter was set to $\sim$ 2 cm by using an iris at the exit of
the lamp house.

The input beam is diffracted by the grating, and the camera section
consisting of lenses and a near-infrared detector (320 $\times$ 256
InGaAs array) is scanned to capture the diffracted beam. The output slit
of the monochromator is thus re-imaged on the detector. Since the
detector exhibits some sensitivity at visible wavelengths, a visible
blocking filter which is transparent at wavelengths longer than 0.75
$\mu$m is inserted after the monochromator to reduce contamination of
visible light from higher orders.

\begin{figure}
\begin{center}
 \includegraphics[height=8cm,angle=-90,keepaspectratio]{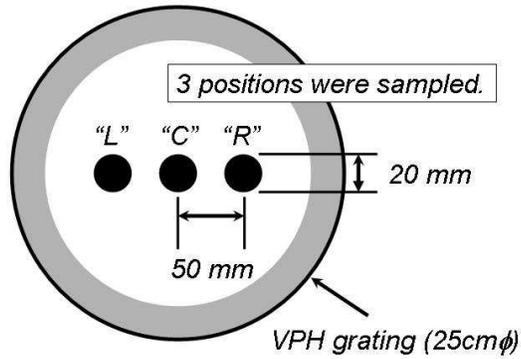}
 \caption{Schematic view of the sampled positions on the VPH grating
 (referred to as ``L'', ``C'', and ``R''). The region near the edge
 indicated in grey is occupied by the support structure of the grating
 mount and is not optically accessible.}\label{pos}
\end{center}
\end{figure}

The basic measurement procedures are as follows: First, the brightness
of the lamp and the wavelength of light exiting from the monochromator
are fixed (the brightness of the lamp is kept constant by a stabilized
power supply during the measurement cycle at a given wavelength), and
the total intensity included in the image of the slit is measured
without the VPH grating. Then, the VPH grating is inserted at an angle
to the optical axis, and the intensities of the zero and first order
($+1$) diffracted light are measured. The diffraction angle is also
recorded. Next, the grating is set at a different incident angle and the
intensities of the diffracted light and diffraction angles are measured
again. After these measurements are repeated for all the incident angles
of interest, a different wavelength is chosen and the same sequence is
repeated. The brightness of the lamp can be changed when moving from one
wavelength to another: a higher brightness is used at shorter
wavelengths because the system throughput is lower. After all the
incident angles and wavelengths are scanned, the grating position is
moved and fixed on the rotation stage so that a different position of
the grating aperture is illuminated. We sampled three positions as
shown in Fig. \ref{pos} (labelled as ``L'', ``C'', and ``R'').
Since each position of the grating illuminated is mechanically located
at the centre of the rotation stage, any aiming errors of illumination
angles to the grating are expected to be equally small at all the
measured positions.

\section{Results and discussions}

\begin{figure}
\begin{center}
 \includegraphics[height=8cm,angle=-90,keepaspectratio]{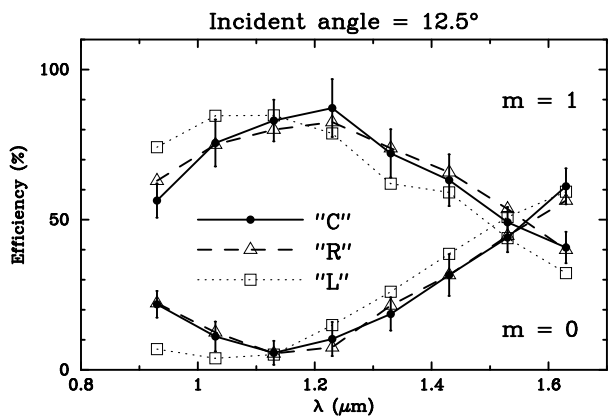}
 \caption{Diffraction efficiency measured for an incident angle of
 12.5$^{\circ}$. Throughputs measured at the centre are shown by solid
 circles (and solid line), and those at the other two positions are
 indicated by open triangle (and dashed line) and open square (and
 dotted line). The upper three lines are for the $m=+1$ order diffracted
 light and the lower ones are for the 0th order light.} \label{eff125}

\end{center}
\end{figure}

\begin{figure}
\begin{center}
 \includegraphics[height=8cm,angle=-90,keepaspectratio]{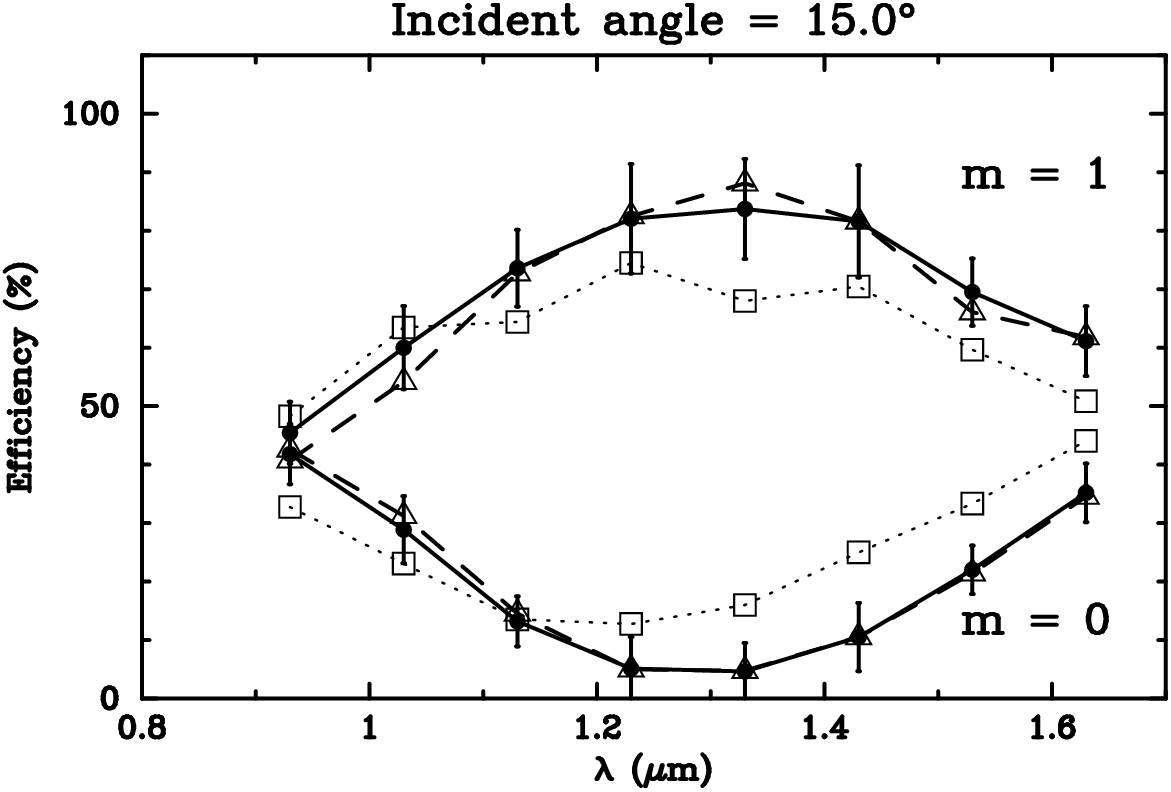}
 \caption{Same for Fig. \ref{eff125}, but for an incident angle of
 15$^{\circ}$.}  \label{eff150}
 \includegraphics[height=8cm,angle=-90,keepaspectratio]{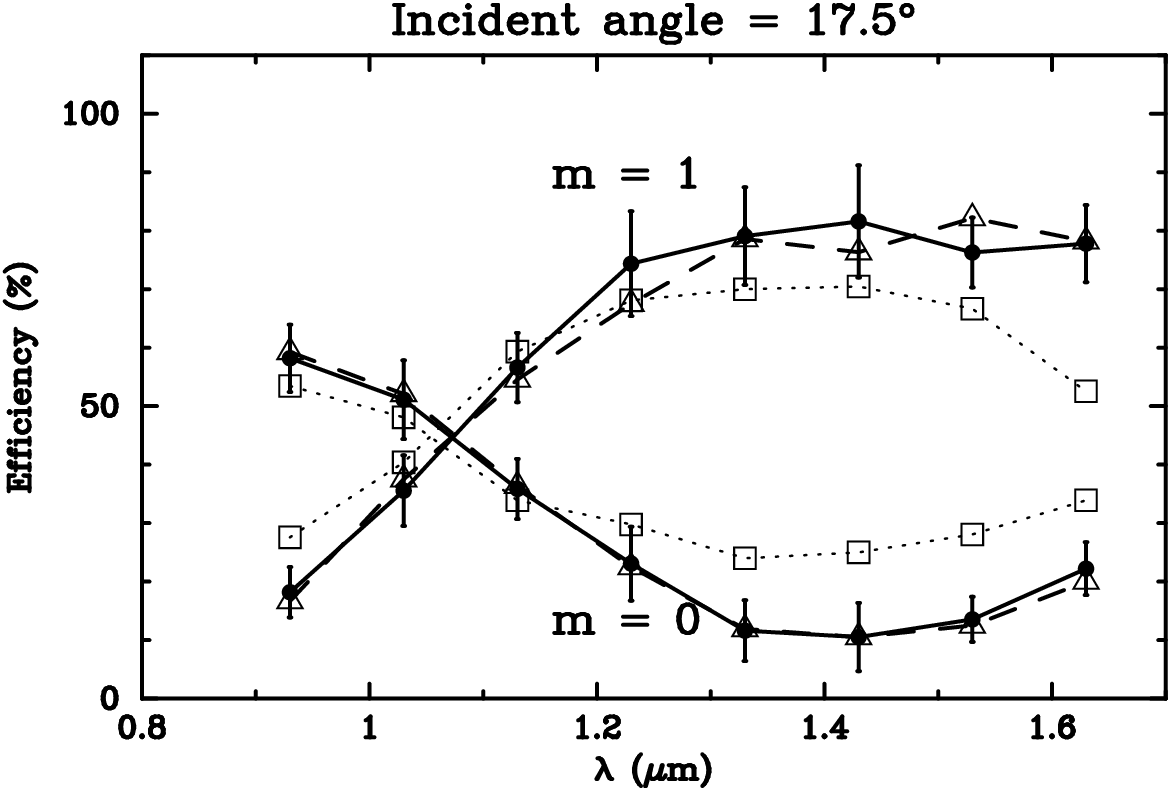}
 \caption{Same for Fig. \ref{eff125}, but for an incident angle of
 17.5$^{\circ}$.}  \label{eff175}
\end{center}
\end{figure}

In Fig. \ref{eff125}, \ref{eff150}, and \ref{eff175}, diffraction
efficiency measured at the three positions is plotted against wavelength
for incident angles of 12.5$^{\circ}$, 15.0$^{\circ}$, and
17.5$^{\circ}$, respectively. Random errors are dominated by
fluctuations of the bias level of the detector on short timescales
($\sim 0.1 - 1$ sec) and the error bars are calculated from a typical
value of these fluctuations (for clarity error bars are shown only for
the data at the position ``C''). The measurements suggest that,
independently of incident angle, the throughputs at the position ``L''
tend to be at variance with the others; the difference in throughput can
be $\sim$ 10 \%. The differences are seen in the measurements both of
the zero and first order diffractions in a mutually compensating manner;
when the throughput of the first order diffraction is lower, that of the
zero order is higher, and vice versa. Also, on the whole, the peak of
the throughput appears to shift towards shorter wavelengths (this is
clearest at an incident angle of 12.5$^{\circ}$). These suggest that the
differences originate in the diffraction process, and are not due
primarily to some additional absorption or reflection at that position.

\begin{figure}
\begin{center}
 \includegraphics[height=8cm,angle=-90,keepaspectratio]{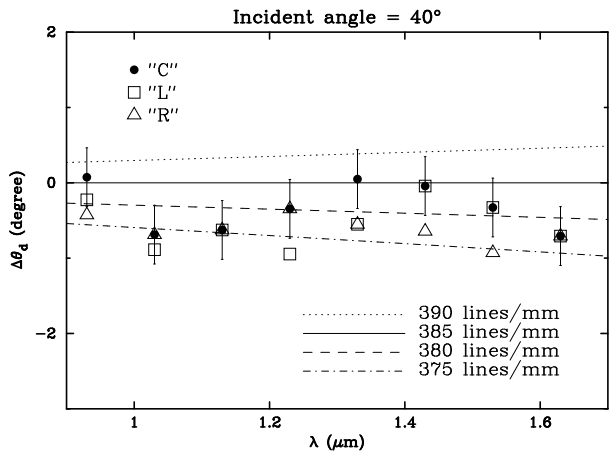}
 \caption{Relationship between diffraction angle and wavelength.
 Difference of diffraction angle from that predicted for a line density
 of 385 lines/mm (solid line; this is the specification of the VPH
 grating) and an incident angle of 40$^{\circ}$ are plotted against
 wavelength. The symbols have the same meanings as in the previous
 figures. Dot-dashed, dashed, and dotted line shows the relationship
 expected for 375, 380, and 390 lines/mm, respectively.}
 \label{dispersion}
\end{center}
\end{figure}

Measurements of diffraction angles are given in Fig. \ref{dispersion}.
The difference of diffraction angle from that predicted for a line
density of 385 lines/mm (the nominal line density of the VPH grating)
and for an incident angle of 40$^{\circ}$ is plotted against wavelength.
The predicted relationship for line density of 375, 380, and 390
lines/mm is also shown for comparison. The data points show the actual
measurements at the three positions. The error is typically
$\sim$0.4$^{\circ}$, which is estimated from a comparison of the
diffraction angles measured at the position ``C'' with those measured
mechanically afterwards for the same incident angle (see \cite{tamura}).
Again, for clarity, the error bars are shown only for the data points at
the position ``C''. A large incident angle was chosen for this test
because the greater the incident angle is, the more sensitive are the
results to changes in line density. While the data points show some
scatter around the predicted relationships, this plot indicates that the
diffraction angles measured at one position exhibit no clear discrepancy
from the others and that the spectral dispersion is nearly independent
of position.
The scatter of the data points in Fig. \ref{dispersion} is probably due
to measurement errors and thus the line density is likely to be rather
uniform across the grating. Line density is normally well controlled in
the manufacturing process, with a typical accuracy of $\sim$ 1 line/mm
over a grating aperture. Nevertheless, it is worth treating the range
between 375 lines/mm and 385 lines/mm, where the data points are
scattered, as an upper limit for the variation of the line density in
the following analysis.

\begin{figure}
\begin{center}
 \includegraphics[height=8cm,angle=-90,keepaspectratio]{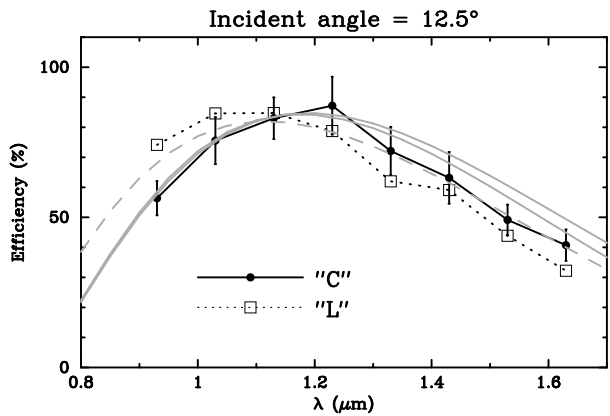}
 \caption{The throughput curves measured at the positions ``C'' and
 ``L'' for an incident angle of 12.5$^{\circ}$ are plotted; these are
 the same as shown in Fig. \ref{eff125}. They are compared with
 predictions from coupled-wave analysis, which are indicated by grey
 smooth (solid and dashed) curves. Grey solid lines indicate the
 predictions for line densities of 375 and 385 lines/mm. In these
 calculations, a refractive index modulation amplitude of 0.05 and a
 dichromated gelatin depth of 12 $\mu$m are assumed. 15\% energy loss is
 also considered (this is expected to be explained by a combination of
 internal absorption and surface reflection). The grey dashed line shows
 the model prediction where a dichromated gelatin depth of 11 $\mu$m is
 assumed instead of 12 $\mu$m. 3 \% additional energy loss (i.e., 18 \%
 in total) presumably caused by more internal absorption is also
 inferred. The same refractive index modulation amplitude is adopted
 (0.05) and the line density is 385 lines/mm.
 It should be mentioned that a nearly identical curve to the grey line
 can be obtained with coupled wave analysis by adopting a refractive
 index modulation amplitude of 0.045 instead of changing the gelatin
 thickness.}  \label{eff125mod}
\end{center}
\end{figure}

\begin{figure}
\begin{center}
 \includegraphics[height=8cm,angle=-90,keepaspectratio]{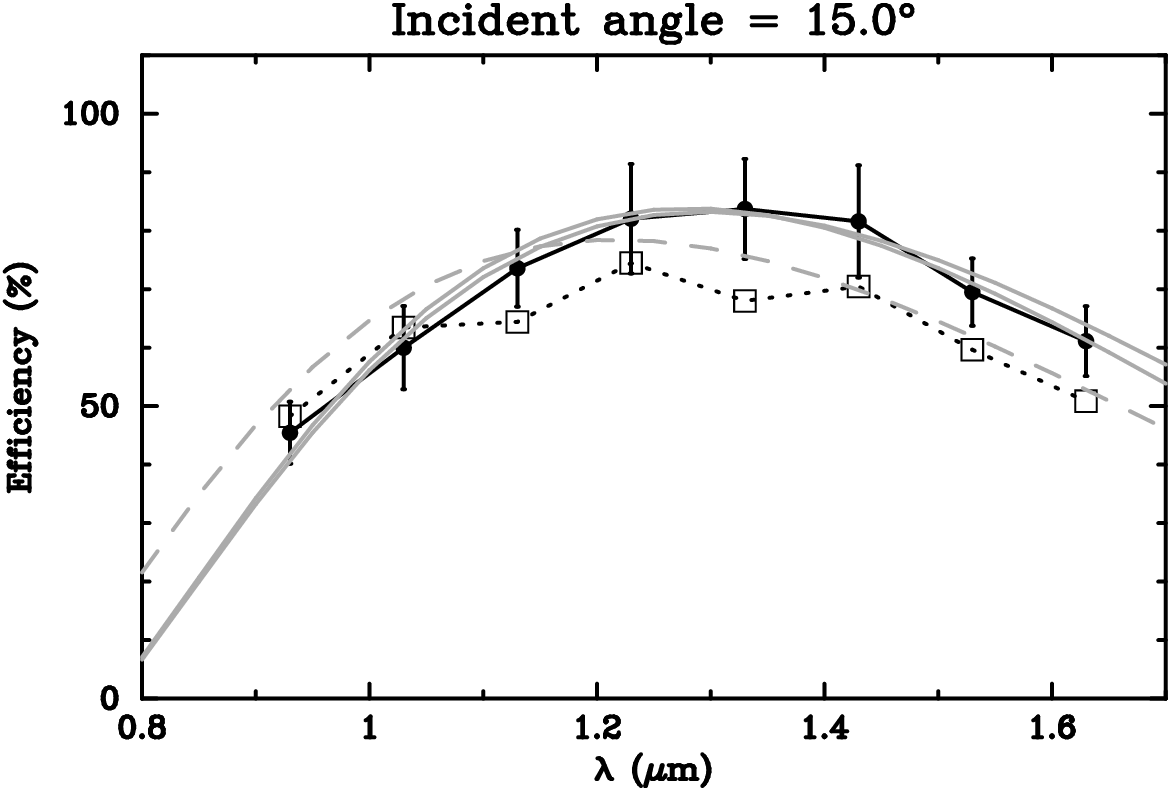}
 \caption{Same as Fig. \ref{eff125mod}, but for an incident angle of
 15.0$^{\circ}$.}
 \label{eff150mod}
 \includegraphics[height=8cm,angle=-90,keepaspectratio]{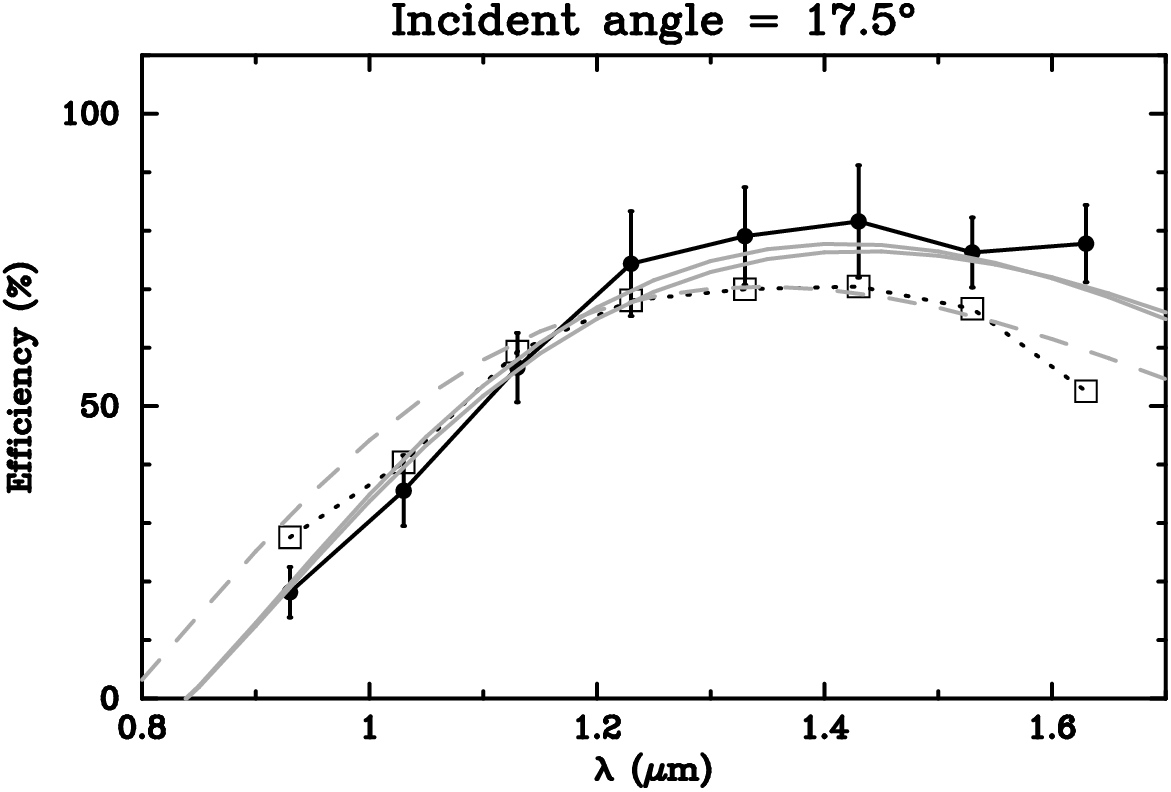}
 \caption{Same as Fig. \ref{eff125mod}, but for the incident angle of
 17.5$^{\circ}$.}
 \label{eff175mod}
\end{center}
\end{figure}

Now we try to identify which parameters vary across the grating, by
comparing the measured diffraction efficiency with the predictions from
coupled-wave analysis (\cite{kogelnik}). In Fig. \ref{eff125mod},
\ref{eff150mod}, and \ref{eff175mod}, the throughput curves measured at
the positions ``C'' and ``L'' for incident angles of 12.5$^{\circ}$,
15.0$^{\circ}$, and 17.5$^{\circ}$ are plotted, respectively, in the
same way as for Fig. \ref{eff125}, \ref{eff150}, and \ref{eff175}. Note
that the throughput curve measured at the position ``R'' is consistent
with that at the position ``C'' and is therefore not shown here for
clarity. Overplotted are three model predictions which are described by
grey smooth (solid and dashed) curves. Two grey solid lines show the
predictions for different line densities; 375 and 385 lines/mm. In
calculating these theoretical predictions, a refractive index modulation
amplitude of 0.05 and a dichromated gelatin depth of 12 $\mu$m are
assumed. Also, 15\% energy loss (presumed to be caused by a combination
of internal absorption and surface reflection) is assumed at all the
wavelengths, to better fit the predictions to the measurements. It
should be emphasized that these assumptions are used in calculating all
the grey solid curves plotted in Fig.  \ref{eff125mod}, \ref{eff150mod},
and \ref{eff175mod}. The energy lost by internal absorption in the
dichromated gelatin layer is estimated to be $\leq$ 1\% below 1.8 $\mu$m
(e.g., \cite{barden1}) and those by combined surface reflections at
boundaries between glass and air are $\sim$ 10\%. Our data may therefore
indicate a larger internal absorption. There may also be some energy
loss due to surface reflections at boundaries between gelatin and glass,
although this is expected to be very small because the refractive
indices are similar.

The fact that the predictions for line densities of 375 and 385 lines/mm
lie close together indicates that even if the line density varies across
the grating at this level (as suggested by Fig.  \ref{dispersion}), the
discrepancy of the throughput curve at the position ``L'' cannot be
fully explained. On the other hand, the grey dashed line, fitting better
to the measurements at the position ``L'', shows the model prediction
where a dichromated gelatin depth of 11 $\mu$m is assumed, instead of 12
$\mu$m. Furthermore, 3\% additional energy loss (hence 18\% in total) is
assumed. The refractive index modulation amplitude is the same as before
(0.05) and the line density is 385 lines/mm.
Nearly identical model predictions to those described with the grey
lines in Fig. \ref{eff125mod}, \ref{eff150mod} and \ref{eff175mod} can
be obtained by adopting a refractive index modulation amplitude of 0.045
instead of changing the gelatin thickness. A combination of small
variations in the gelatin thickness and/or refractive index modulation
amplitude, and the energy loss due to internal absorption and/or surface
reflection are therefore equally likely candidates for the throughput
variation, at least for this grating. Variations of the gelatin
thickness across a VPH grating by $\sim$ 1 $\mu$m have been suggested by
direct measurements on a 14 cm $\times$ 15 cm grating with a line
density of 850 lines/mm (\cite{blanche}). However, since refractive
index modulation amplitude is proportional to exposure, an uneven
exposure within the grating aperture is also a possible cause of the
spatial variation.

\section{Summary \& Conclusion}

In this paper, we report measurements of diffraction efficiency and
angular dispersion of a large format ($\sim$ 25 cm diameter) VPH grating
optimized for near-infrared wavelengths (0.9 $\sim$ 1.8 $\mu$m). The aim
of this experiment is to see whether optical characteristics vary
significantly across the grating. We investigated three positions in the
grating aperture with a separation of 5 cm between each, using a 2 cm
diameter beam to illuminate the grating. At each position, throughput
and diffraction angles were measured at several wavelengths. Our data
indicate that whilst the line density is nearly constant at the three
positions ($\leq$ 3\% variation), the throughput at one position can be
different by $\sim$ 10\% due mainly to a small shift of the throughput
peak towards shorter wavelengths. Comparing the data with predictions
from coupled-wave analysis, we find that this throughput variation can
be explained by a combination of small variations in gelatin depth
and/or refractive index modulation amplitude and amount of energy lost
by internal absorption and/or surface reflection.
If a grating is used at a pupil plane of the spectrograph, such
variations will be averaged out for all points in the field of view.
However, if it is used in a non-pupil plane, careful specification of
tolerances would be required to obtain a system with good astronomical
performance.

In the near future, we intend to perform similar experiments to those
reported here for several VPH gratings to see variations across a
selection of gratings with increasing sampling points in the grating
aperture. We will also conduct these experiments after exposing the
gratings to vacuum and cryogenic temperatures. These results will be
reported in a forthcoming paper.

\section*{Acknowledgements}
We thank colleagues in Durham for their assistance with this work,
particularly Peter Luke, John Bate, and the members of the mechanical
workshop. We also thank the anonymous reviewer for his/her comments
which were very useful to improving this paper. This work was funded by
PPARC Rolling Grant PPA/G/O/2000/00485.

\end{document}